\documentclass[final,1p,times]{elsarticle}

\usepackage{graphicx}
\usepackage{amssymb} 
\usepackage{amsthm} 
\usepackage{wrapfig}

\journal{Nuclear Physics A} 

\begin{document}

\begin{frontmatter} 

\title{Beam Energy Dependence of First and Higher-Order Flow Harmonics from the STAR Experiment at RHIC}

\author  {Yadav Pandit  (for the STAR Collaboration)}
\address { Department of Physics, Kent State University, Kent, OH 44242}

\begin{abstract} 
In these proceedings, we present STAR measurements of directed flow, $v_{1}$, for pions, protons and antiprotons between $\sqrt{s_{NN}} = 7.7$ GeV and 200 GeV.  A striking observation is that the $v_{1}$ slope for net protons as a function of normalized rapidity $y'$ ($y/y_{\rm beam}$), $F = dv_{1}/dy'$, which is an estimate of the directed flow contribution from baryon number transported to the midrapidity region, changes sign twice within the lower part of this energy range. We also present the measured beam energy dependence of triangular flow $v_{3}$, and the $n=1$ to $n=5$ flow harmonics for charged particles at 200 GeV.
\end{abstract} 

\end{frontmatter} 


\section{Introduction}
The study of azimuthal anisotropy, based on Fourier coefficients, is widely recognized as an important tool to probe the hot, dense matter created in heavy ion collisions. In non-central collisions, the initial volume of the interacting system is anisotropic in coordinate space. Due to multiple interactions, this anisotropy is transferred to momentum space and is then quantified via the flow harmonics $v_{n}$~\cite{methodPaper}, defined by
\begin{equation}
v_n = \langle \cos n( \phi-\Psi_R ) \rangle, 
\end{equation}
where $\phi$ denotes the azimuthal angle of an outgoing particle, $\Psi_R$ is the azimuth of the reaction plane, and $n$ denotes the harmonic order.  
Higher order flow coefficients may provide new insights into initial-state fluctuations, and the subsequent evolution of the collision system~\cite{geoFluct1}. 

\section{Methods and Analysis}
The STAR detector provides full azimuthal coverage near midrapidity $|\eta|< 1.0$ and is an ideal detector for flow measurement.  The Time Projection Chamber (TPC) is used as the main detector for charged particle tracking at midrapidity.  Protons and antiprotons up to 2.8 GeV/$c$ and $\pi^\pm$ and $K^\pm$ up to 1.6 GeV/$c$ in transverse momentum are identified based on specific energy loss in the TPC and time-of-flight information from a multi-gap resistive plate chamber barrel.  We use charged particles in the TPC  for event plane reconstruction for higher order flow harmonic measurement~\cite{yadav}. To avoid self-correlations, we calculate the event plane in two separate pseudorapidity hemispheres, with a gap of 0.1 units between them. For directed flow analysis in the range $\sqrt{s_{NN}} = 7.7$ to 39 GeV, two Beam-Beam Counters (BBC), covering $3.3 < |\eta| < 5.0$ were used to reconstruct the first-order event plane. The pseudorapidity gap between BBC and TPC suppresses non-flow effects. 

\section{Results}
\subsection{Directed flow}

\begin{wrapfigure}{r}{0.5\textwidth}
  \begin{center}
    \includegraphics[width=0.48\textwidth]{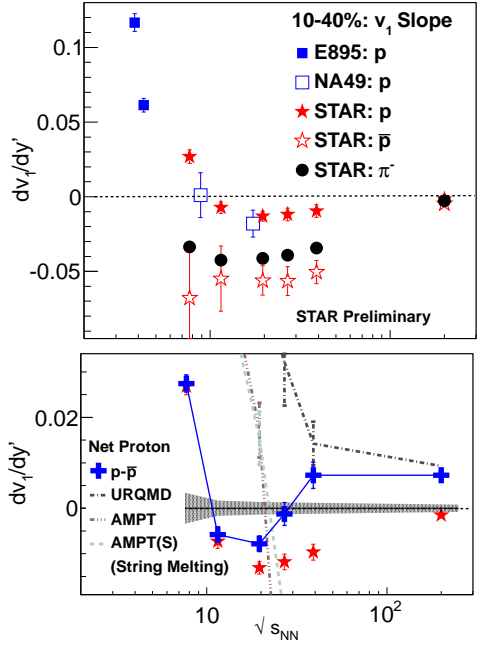}
  \end{center}
  \caption{Directed flow slope ($dv_{1}/dy'$) near midrapidity as a function of beam energy for mid-central(10-40\%) Au+Au collisions, where the primed quantity $y'$ refers to normalized rapidity $y/y_{\rm beam}$.  The upper panel reports slopes for protons, antiprotons and pions, including measurements by prior experiments.  The lower panel shows STAR's measurement of this same slope for net protons, which is a representation of the signal for initial-state baryon number transported to midrapidity, along with corresponding predictions from the UrQMD~\cite{UrQMD} and AMPT~\cite{AMPT} models. The systematic uncertainty on the net-proton measurements is shown as a shaded band centered on $dv_1/dy' =0$.}
\label{directedFlow}
\end{wrapfigure}
Directed flow as a function of beam energy is considered to be sensitive to a first-order phase transition  and may provide a possible signature of a softening of the Equation of State~\cite{Stoecker}. Figure~\ref{directedFlow} shows directed flow slope ($dv_{1}/dy'$) near midrapidity as a function of beam energy for 10-40\% central Au+Au collisions.  In these mid-central collisions, the proton $v_{1}$ slope in the midrapidity region changes sign from positive to negative between 7.7 and 11.5 GeV and remains small but negative up to 200 GeV, while the slope for $\pi^\pm$, $K^\pm$ and antiprotons is always negative. The $v_{1}$ slope for net protons (where produced protons are removed) is calculated  based on relation $F = r F_{\bar{p}} + (1-r) F_{\rm transp}$, where $r$ is the observed ratio of antiprotons to protons among the analyzed tracks.  The net proton $v_{1}$ slope displays a prominent dip in the vicinity of $\sqrt {s_{NN}} =$10--20 GeV, changing sign twice within the studied energy range. This result is qualitatively different from UrQMD and AMPT transport models, which both predict a monotonic trend throughout $\sqrt {s_{NN}} = $7.7 to 200 GeV. However, the predictions of these two models strongly differ from each other. Further clarification of possible explanations unrelated to the Equation of State calls for specific additional experimental and theoretical studies.

\subsection{Triangular flow}
Theoretical studies suggest that $v_{3}$ is more sensitive to viscosity than $v_{2}$, because the finer details present in the higher harmonics are smoothed more by viscous effects~\cite{geoFluct1}. Figure 2 shows the beam energy dependence of triangular flow near mid-rapidity as a function of transverse momentum for beam energies from 7.7 to 200 GeV in Au+Au collisions. It is observed that the centrality dependence of triangular flow is weak over all the energies studied.  This might be due to either the centrality dependence of the spatial ellipticity and triangularity being different, and/or viscous effects being different. The lower panel reports ratios with $v_{3}$ at 200 GeV in the denominator and at  lower $p_{T}$, we observe beam energy dependence.

\begin{figure}[htbp]
\begin{center}
 \includegraphics[width=1.0 \textwidth]{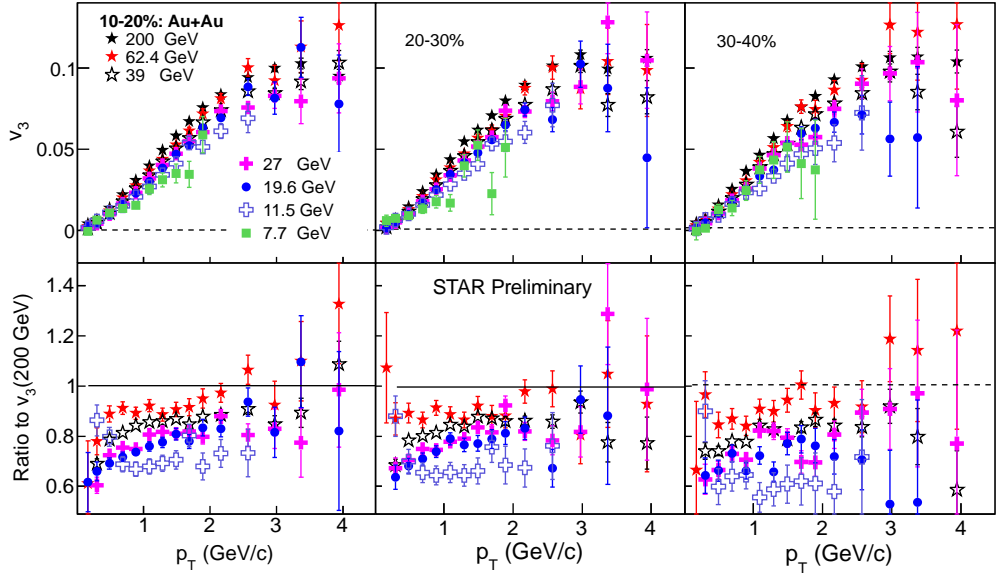}
\end{center}
\caption{ Triangular flow $v_3$ as a function of transverse momentum in Au+Au collisions from 7.7 to 200 GeV. These measurements were performed with respect to the third-order event plane reconstructed in the TPC using the $\eta$ sub-event method. }
\label{fig:TriangularFlow}
\end{figure}

\subsection{Higher-order flow harmonics}
\begin{figure}[htbp]
\begin{center}
 \includegraphics[width=1.0 \textwidth]{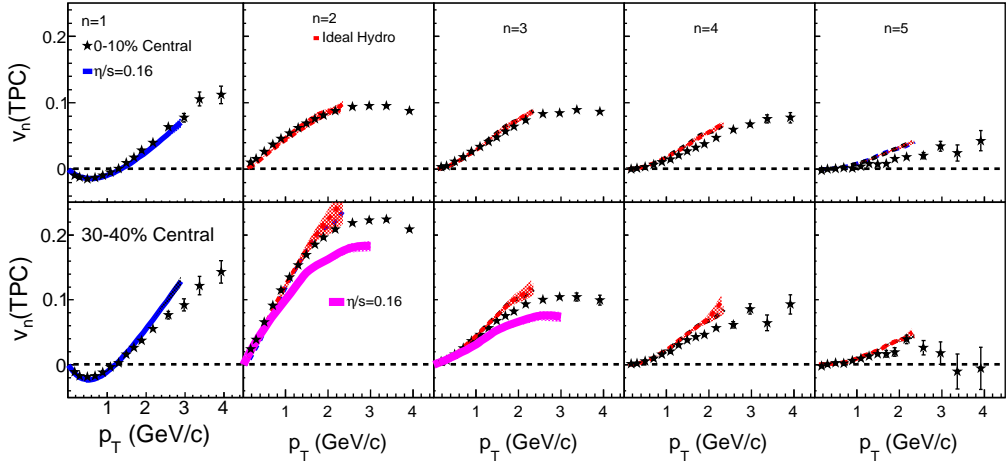}
\end{center}
\caption{ Flow harmonics $v_{n}$ for $n=1$ to $n=5$ as a function of transverse momentum in 200 GeV Au+Au collisions. These measurements were done with respect to the $n$th-order event plane reconstructed in the TPC using the $\eta$ sub-event method.  The model curves are from Refs.~\cite{Luzum,Gardim,bshake}. }
\label{fig:harmony}
\end{figure}

The $v_{n}$ of all charged hadrons, for $n =1,\, 2,\, 3,\, 4$ and 5 are measured using the event plane method as a function of $p_{T}$ at various centralities, as shown in Fig. 3.  The data for 0--10\% centrality are shown in the upper panels, and for intermediate centrality (30--40\%) in the lower panels. The first harmonic coefficient shown in this figure is a rapidity-even signal related to the dipole asymmetry in the initial geometry, and is measured with a modified event-plane method~\cite{Luzum} that suppresses the conventional rapidity-odd directed flow signal, and also corrects for momentum conservation effects. All $v_{n}$ measurements show an increasing trend as a function of $p_{T}$.  Except for $v_{2}$, we observe very weak centrality dependence.  The initial overlap geometry, which makes the dominant contribution to the elliptic flow, changes from central to peripheral collisions, giving rise to the strong centrality dependence of $v_{2}$. The centrality dependence of the geometry fluctuations are not yet understood. The weak centrality dependence of the flow coefficients which are believed to originate from geometry fluctuations indicates that geometry fluctuations weakly depend on centrality.  The solid lines in Fig. 3 are model predictions. The model prediction for $n=1$ are from Ref.~\cite{Luzum}, where the ratio of shear viscosity to entropy density is $\eta/s =
0.16$, whereas for other harmonics, the model predictions are for ideal hydrodynamics~\cite{Gardim}. For $v_{2}$ and $v_{3}$, we also include the curves from Ref.~\cite{bshake}, where viscous hydrodynamical model was assumed with $\eta/s = 0.16$.   

\section{Summary}
For mid-central Au+Au collisions, the slope for proton directed flow changes sign from positive to negative between 7.7 and 11.5 GeV and remains small 
but negative up to 200 GeV. The $v_{1}$ slope for net protons shows non-monotonic behavior at low beam energies; it is negative at 11.5 and 19.6 GeV and 
positive at all other energies. UrQMD and AMPT transport models do not show even qualitatively similar behavior.  More input from theory is necessary to 
interpret these measurements, and additional experimental statistics would permit a study of centrality dependence. Triangular flow as a function of $p_{T}$ 
is reported for Au+Au in the same energy scan region (7.7- 39 GeV), and is compared with 62.4 and 200 GeV Au+Au collisions.  We also report the 
centrality dependence of $v_{3}$. All flow harmonics ($n=$1--5) have been measured as a function of $p_{T}$. These measurements provide significant 
inputs for understanding both initial fluctuations and transport properties.

\end{document}